\documentclass{desyproc}

\usepackage{slashed}
\usepackage{bbold}
\usepackage{wasysym}
\usepackage{cite}
\usepackage[russian,english]{babel}

\begin{document}
\title{The quantum vacuum in electromagnetic fields: From the Heisenberg-Euler effective action to vacuum birefringence}

\author{{\slshape Felix Karbstein$^{1,2}$}\\[1ex]
$^1$Theoretisch-Physikalisches Institut, Abbe Center of Photonics, Friedrich-Schiller-Universit\"at Jena, Max-Wien-Platz 1, D-07743 Jena,  Germany \\
$^2$Helmholtz Institut Jena, Fr\"obelstieg 3, D-07743 Jena, Germany}

\contribID{xy}

\confID{999}  
\desyproc{DESY-PROC-2099-01}
\doi  

\maketitle

\begin{abstract}
 The focus of these lectures is on the quantum vacuum subjected to classical electromagnetic fields.
 To this end we explicitly derive the renowned Heisenberg-Euler effective action in constant electromagnetic fields in a rather pedagogical and easy to conceive way.
 As an application, we use it to study vacuum birefringence constituting one of the most promising optical signatures of quantum vacuum nonlinearity.
\end{abstract}

\section{Introduction}

These lecture notes are about the quantum vacuum in strong electromagnetic fields. 
They are intended to provide interested students having basic knowledge in quantum field theory with an introduction to this topic.
To this end it is instructive to first recall the classical understanding of the vacuum:
In a classical field theory context, the vacuum can be defined as ``the absence of anything,''
and can be considered as a kind of absolute ground state, not featuring any excitations, etc.

When adding classical electromagnetic fields to this vacuum, we arrive at the classical field theory of electrodynamics, which can be defined by the following Lagrangian density,
\begin{equation}
 {\cal L}_\text{ED}=-\frac{1}{4}F^{\phantom{\mu}}_{\text{cl.}\mu\nu}F_\text{cl.}^{\mu\nu}-j_\mu A_\text{cl.}^\mu \,,
\end{equation}
where $F_\text{cl.}^{\mu\nu}=\partial^\mu A_\text{cl.}^\nu-\partial^\nu A_\text{cl.}^\mu$ is the field strength tensor, and $j^\mu$ a current sourcing the electromagnetic four-potential $A_\text{cl.}^\mu$.
In these lectures, we are only interested in the vacuum subjected to electromagnetic fields far away from any classical sources. Correspondingly, we set $j_\mu=0$.

The equations of motion for a Lagrangian ${\cal L}(A^\mu,F^{\mu\nu})$ are given by the following Euler-Lagrange equations,
\begin{equation}
 \frac{\partial{\cal L}}{\partial A_\nu}-2\partial_\mu\frac{\partial{\cal L}}{\partial F_{\mu\nu}}=0 \,, \label{eq:EulerLagrange}
\end{equation}
such that for ${\cal L}(A^\mu,F^{\mu\nu})\to{\cal L}_\text{ED}(A_\text{cl.}^\mu,F_\text{cl.}^{\mu\nu})|_{j=0}$ we obtain the free Maxwell equations, $\partial_\mu F_\text{cl.}^{\mu\nu}=0$.
Obviously, these homogeneous linear differential equations have the feature that any combination $F_\text{cl.}^{\mu\nu}=\sum_i F_{\text{cl.} i}^{\mu\nu}$ of solutions $F_{\text{cl.}i}^{\mu\nu}=(\partial^\mu g^{\nu}_{\ \alpha}-\partial^\nu g^{\mu}_{\ \alpha})A_{\text{cl.}i}^\alpha$ is also a solution, which implies the superposition principle to hold for electromagnetic fields in the {\it vacuum of classical electrodynamics}.

In a next step we turn to the notion of the vacuum in a quantum field theory (QFT) context, also referred to as the quantum vacuum.
The quantum vacuum is not empty, but rather permeated by fluctuations of the particle degrees of freedom of the theory under consideration (in quantum electrodynamics: electrons, positrons and photons) in so-called virtual processes.
In contrast to real particles, virtual particles are typically not on the mass shell, i.e., do not fulfill the relativistic energy-momentum relation.
These virtual processes can be excited by external influences, e.g., electromagnetic fields \cite{Heisenberg:1935qt}, or by imposing physical boundary conditions \cite{Casimir:dh}.
Aiming at unveiling the quantum nature of the vacuum in an experiment (Gedankenexperiment or actual experiment), the basic idea is to look for potential imprints of the fluctuating fields in the measured response.

In these lectures we limit ourselves to the vacuum of quantum electrodynamics (QED), defined by the following Lagrangian,
\begin{equation}
 {\cal L}_\text{QED}=-\frac{1}{4}F_{\mu\nu}F^{\mu\nu}+\bar\psi({\rm i}\slashed{D}-m)\psi \,, \label{eq:L_QED}
\end{equation}
with $\bar\psi=\psi^\dag\gamma^0$, $\slashed{D}=\gamma^\mu D_\mu$, $D_\mu=\partial_\mu-{\rm i}eA_\mu$ and $F^{\mu\nu}=\frac{\rm i}{e}[D^\mu,D^\nu]$, where $e$ is the electron charge. The latter term mediates the coupling of the four-component Dirac spinors $\psi$, describing spin-$\frac{1}{2}$ fermions of mass $m$ to the photon field $A^\mu$. Note that $\psi$ accounts for both electron and positrons degrees of freedom.
The defining property of the Dirac gamma matrices $\gamma^\mu$ is $\{\gamma^\mu ,\gamma^\nu\}=-2g^{\mu\nu}\mathbb{1}$, and our metric convention is $g_{\mu \nu}=\mathrm{diag}(-1,1,1,1)$.
We use the Heaviside-Lorentz System with $c=\hbar=1$. 
A slightly different representation of Eq.~\eqref{eq:L_QED} is
\begin{equation}
 {\cal L}_\text{QED}=-\frac{1}{4}F_{\mu\nu}F^{\mu\nu}+\bar\psi({\rm i}\slashed{\partial}-m)\psi+eA^\mu\bar\psi \gamma_\mu\psi \,, \label{eq:L_QED_v2}
\end{equation}
\vspace*{-3.5mm}

\hspace*{2mm}\centerline{\includegraphics[height=1.02cm]{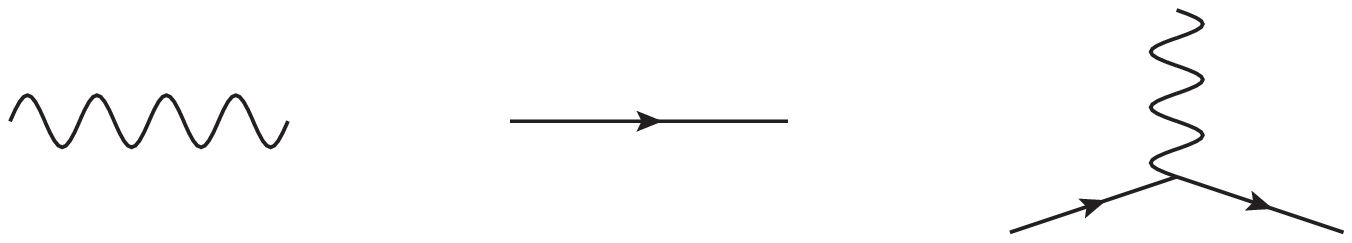}}\vspace{0.8mm}
where we split ${\cal L}_\text{QED}$ into three terms: The first one is the Maxwell term, describing the dynamics of free electromagnetic fields quantized in terms of photons. The second one represents freely propagating electrons and positrons. 
Finally, the third one accounts for the interaction between the fermions and the photons, whose coupling is parameterized by $e$.

In addition, we provide the graphical representation of these terms in the diagrammatic language of Feynman diagrams.
From these building blocks all possible QED processes can be constructed, following the simple recipe that {\it every diagram that can be constructed by assembling the above elementary building blocks can happen} (if it is not vanishing due to symmetries, kinematic restrictions, etc.).
Generically, this gives rise to loop diagrams.
In fact, for any given physical process specified by fixing the external legs, characterizing the incident and outgoing real particles, there are infinitely many loop diagrams.
As they do not involve any real particles (in QED: in- and outgoing photons, electrons and positrons) pure vacuum-type diagrams correspond to Feynman diagrams without external lines.
Favorably for us, in the regime of validity of perturbative QED, loop diagrams are more and more suppressed with increasing loop order, as each additional loop results in a parametric suppression by a factor of $\alpha=\frac{e^2}{4\pi}$.

\section{Towards the Heisenberg-Euler effective Lagrangian}

Let us now turn to QED in an external classical background field $A^\mu_\text{cl.}$ and for the moment completely ignore quantized dynamical photons.
This ad hoc assumption can actually be justified by the fact that for the effect we are interested in here, the leading order contribution is insensitive to dynamical photons. 
Their presence only becomes important at higher loop order; cf. also below.

Hence, from now on we focus on the Lagrangian~\eqref{eq:L_QED} with the replacement $A_\mu\to A_{\text{cl.}\mu}$.
The Lagrangian and Feynman rules for this theory are given by
\begin{equation}
 {\cal L}=-\frac{1}{4}F^{\phantom{\mu}}_{\text{cl.}\mu\nu}F_\text{cl.}^{\mu\nu}+\bar\psi({\rm i}\slashed{\partial}-m)\psi+eA_{\text{cl.}\mu}\bar\psi \gamma^\mu\psi \,, \label{eq:L_nQED}
\end{equation}
\vspace*{-3.5mm}

\hspace*{2mm}\centerline{\includegraphics[height=1.1cm]{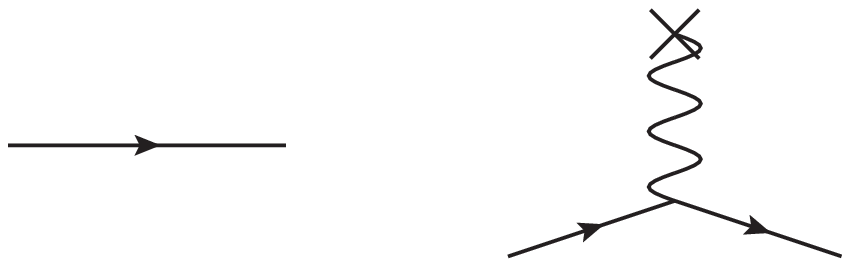}}\vspace{0.8mm}
i.e., the only quantum degrees of freedom of this theory are electrons and positrons. In the Feynman rules, the external character of $A_{\text{cl.}\mu}$ is symbolized by the wiggly line ending at a cross.

Aiming at pure vacuum phenomena, we are only interested in effective interactions among electromagnetic fields, and do not consider situations with {\it real} electrons or positrons in the initial and final states.
However, let us emphasize once more that within a quantum field theory we can never prevent the quantum fields constituting the theory under consideration to occur as internal, {\it virtual} states.
In turn, it is immediately clear that all possible connected Feynman diagrams without external electron/positron lines which can be drawn for the theory defined by the Lagrangian~\eqref{eq:L_nQED} can be represented in terms of a single one-loop diagram,

\vspace*{1mm}
\centerline{\includegraphics[height=1cm]{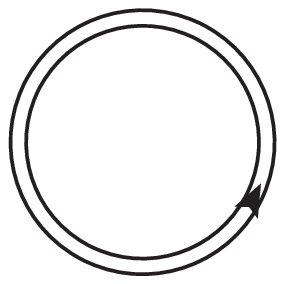}}\vspace{0.8mm}
\noindent with ``dressed'' fermion propagator

\hspace*{2mm}\centerline{\includegraphics[height=1cm]{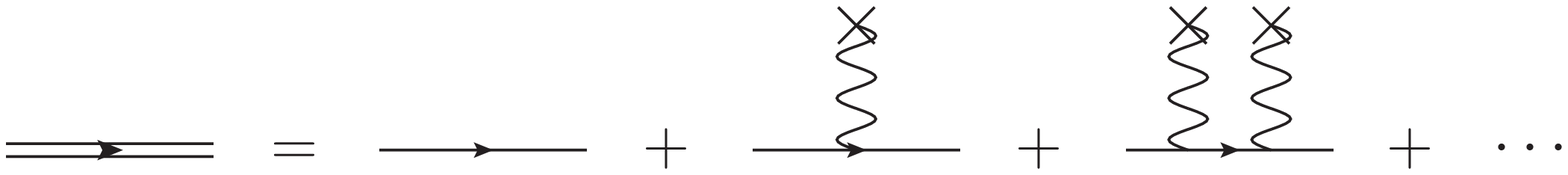}.}\vspace{0.8mm}
In order to evaluate this diagram explicitly, we turn to the partition function $Z$ associated with Eq.~\eqref{eq:L_nQED},
\begin{equation}
 Z[A_\text{cl.}^\mu]={\cal N}\int{\cal D}\bar\psi\int{\cal D}\psi\,{\rm e}^{{\rm i}S }\,,
\end{equation}
where $S=\int{\rm d}^4x\,{\cal L}$ is the corresponding action, and $\cal N$ denotes an overall normalization factor to be fixed below.
As the action is bilinear in the Dirac field $\psi$, the functional integrations can be performed explicitly, providing us with an {\it effective action} $S_\text{eff.}[A_\text{cl.}^\mu]$, which only depends on $A_\text{cl.}^\mu$.
To this end, recall that $\int{\cal D}\bar\psi\int{\cal D}\psi\exp\bigl\{{\rm i}\int{\rm d}^4 x\, \bar\psi M\psi\bigr\}=\det M$.
Introducing the classical Maxwell action $S_\text{MW}=\int{\rm d}^4x\bigl(-\frac{1}{4}F^{\phantom{\mu}}_{\text{cl.}\mu\nu}F_\text{cl.}^{\mu\nu}\bigr)$ and the definition $D_{\text{cl.}\mu}=\partial_\mu-{\rm i}eA_{\text{cl.}\mu}$, we obtain
\begin{align}
 Z[A_\text{cl.}^\mu]&={\cal N}{\rm e}^{{\rm i}S_\text{MW}}\det({\rm i}\slashed{D}_\text{cl.}-m) \nonumber\\
                    &={\cal N}{\rm e}^{{\rm i}[S_\text{MW}-{\rm i}\ln\det({\rm i}\slashed{D}_\text{cl.}-m)]} \nonumber\\
                    &=:{\rm e}^{{\rm i}S_\text{eff.}[A_\text{cl.}^\mu]} \,.
\end{align}
It is convenient to demand that $Z[A_\text{cl.}^\mu=0]=1$, or equivalently $S_\text{eff.}[A_\text{cl.}^\mu=0]=0$, so that the effective action vanishes at zero field.
In turn, we have ${\cal N}=\det^{-1}({\rm i}\slashed{\partial}-m)$, and the effective action can be expressed as
\begin{equation}
 S_\text{eff.}[A_\text{cl.}^\mu]=S_\text{MW}[A_\text{cl.}^\mu]+S^1[A_\text{cl.}^\mu]\,,
\end{equation}
where 
\begin{equation}
 S^1[A_\text{cl.}^\mu]=-{\rm i}\ln\det({\rm i}\slashed{D}_\text{cl.}-m)+{\rm i}\ln\det({\rm i}\slashed{\partial}-m)
\end{equation}
\hspace*{0.81cm}\centerline{\includegraphics[height=1cm]{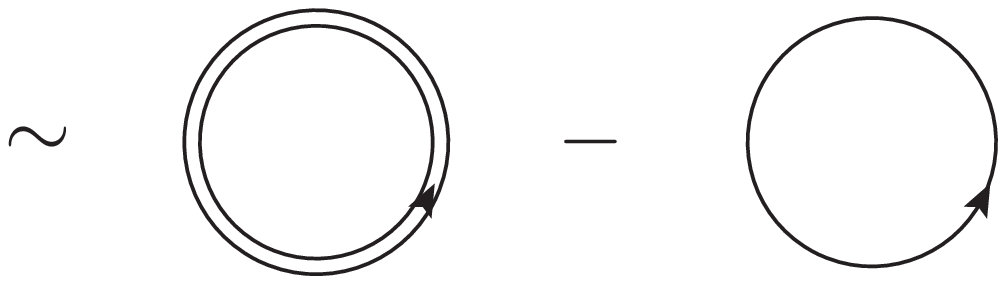}}\vspace{0.8mm}
is the so-called one-loop effective action.
Accounting for the fluctuations of dynamical photons, of course also additional diagrams mediating effective interactions among classical electromagnetic fields arise. As these diagrams have at least one internal photon line, they are of higher-loop order and thus suppressed by additional powers of $\alpha$ as compared to the one-loop result.

In even space-time dimensions we can define a matrix $\gamma_5$ ($\gamma_5={\rm i}\gamma^0\gamma^1\gamma^2\gamma^3$) which fulfills $\{\gamma^\mu,\gamma_5\}=0$ and $(\gamma_5)^2=\mathbb{1}$.
Therewith, $\det({\rm i}\slashed{D}_\text{cl.}-m)=\det[({\rm i}\slashed{D}_\text{cl.}-m)\gamma_5^2]=\det[\gamma_5(-{\rm i}\slashed{D}_\text{cl.}-m)\gamma_5]=\det(-{\rm i}\slashed{D}_\text{cl.}-m)$, such that we can write
\begin{equation}
 S^1[A_\text{cl.}^\mu]=-\frac{\rm i}{2}\ln\det(\slashed{D}_\text{cl.}^2+m^2)+\frac{\rm i}{2}\ln\det(\slashed{\partial}^2+m^2)\,.
\end{equation}
Moreover, note that
\begin{align}
 \slashed{D}_\text{cl.}^2&=D_{\text{cl.}\mu}D_{\text{cl.}\nu}\gamma^\mu\gamma^\nu=D_{\text{cl.}\mu}D_{\text{cl.}\nu}\tfrac{1}{2}
 \bigl(\{\gamma^\mu,\gamma^\nu\}+[\gamma^\mu,\gamma^\nu]\bigr) \nonumber\\
 &=-D_{\text{cl.}}^2-\tfrac{e}{2}\sigma^{\mu\nu}F_{\text{cl.}\mu\nu} \,,
\end{align}
where we employed $\{\gamma^\mu ,\gamma^\nu\}=-2g^{\mu\nu}\mathbb{1}$, $F_\text{cl.}^{\mu\nu}=\frac{\rm i}{e}[D_\text{cl.}^\mu,D_\text{cl.}^\nu]$ (cf. above) and the definition $\sigma^{\mu\nu}\equiv\frac{\rm i}{2}[\gamma^\mu,\gamma^\nu]$,
fulfilling $\frac{1}{2}\{\sigma^{\mu\nu},\sigma^{\alpha\beta}\}=g^{\mu\alpha}g^{\nu\beta}-g^{\mu\beta}g^{\nu\alpha}+{\rm i}\gamma_5\epsilon^{\mu\nu\alpha\beta}$ \cite{Schwinger:1951nm,Dittrich:2000zu}.
Hence, with $\ln\det M={\rm Tr}\ln M$, we have
\begin{equation}
 S^1[A_\text{cl.}^\mu]=-\frac{\rm i}{2}{\rm Tr}\ln(-D_{\text{cl.}}^2+m^2-\tfrac{e}{2}\sigma^{\mu\nu}F_{\text{cl.}\mu\nu})+\frac{\rm i}{2}{\rm Tr}\ln(-\partial^2+m^2)\,.
\end{equation}
The trace ${\rm Tr}\{\ldots\}={\rm tr}_\gamma{\rm tr}_x\{\ldots\}$ is over both spinor indices and coordinate space; we denote the trace over spinor indices by ${\rm tr}_\gamma\{\ldots\}$ and the trace over coordinate space by ${\rm tr}_x\{\ldots\}$.
In the next step we employ the {\it proper time representation} of the logarithm \cite{Schwinger:1951nm},
\begin{equation}
 -\ln M + \ln M_0 = \lim_{\Lambda\to\infty}\int_{-{\rm i}/\Lambda^2}^\infty\frac{{\rm d}T}{T}\bigl({\rm e}^{-{\rm i}MT}-{\rm e}^{-{\rm i}M_0T}\bigr)\,,
\end{equation}
valid for $\Im\{M,M_0\}<0$.
Here, $\Lambda$ (of dimension mass) is a regulator for potentially occurring ultraviolet (UV) divergences.
The precise form of the integral's lower bound is chosen such that if in addition $\Re\{M,M_0\}>0$ we can use a contour rotation ${\rm i}T\to T$ to write
$-\ln M + \ln M_0 = \lim_{\Lambda\to\infty}\int_{1/\Lambda^2}^\infty\frac{{\rm d}T}{T}\bigl({\rm e}^{-MT}-{\rm e}^{-M_0T}\bigr)$.
In turn, we obtain the following proper time representation of the one-loop effective action,
\begin{equation}
 S^1[A_\text{cl.}^\mu]=\frac{\rm i}{2}\int_{-{\rm i}/\Lambda^2}^\infty\frac{{\rm d}T}{T}\,{\rm e}^{-{\rm i}m^2T}\Bigl\{{\rm Tr}\,{\rm e}^{{\rm i}(D_{\text{cl.}}^2+\frac{e}{2}\sigma^{\mu\nu}F_{\text{cl.}\mu\nu})T}-{\rm Tr}\,{\rm e}^{{\rm i}\partial^2 T}\Bigr\}\,, \label{eq:S1A}
\end{equation}
where $\lim_{\Lambda\to\infty}$ is implicitly understood, and to be taken at the very end of the calculation.
Let us emphasize that this expression is still valid for arbitrary fields $A_\text{cl.}^\mu(x)$.
It, or at least the integrand of the proper time integral, i.e., ${\rm Tr}\{\ldots\}$, can be evaluated analytically for several special cases:
\begin{itemize}
 \item A parameter integral representation of $S^1$ is known for {\it constant electromagnetic fields}, $F^{\mu\nu}_\text{cl.}=\text{const.}$, of arbitrary relative orientation of $\vec{E}$ and $\vec{B}$
 \cite{Heisenberg:1935qt,Schwinger:1951nm}; cf. also the explicit derivation presented below. For completeness, note that for this case also the two-loop effective action is known explicitly in terms of a parameter integral representation \cite{Ritus:1975}.
 \item For plane wave {\it null fields} $A_\text{cl.}^\mu(x)\sim\cos(\kappa^\mu x_\mu)$, with $\kappa^2=0$, $\vec{E}\perp\vec{B}$ and $|\vec{E}|=|\vec{B}|$, the effective action vanishes identically, i.e., $S^1=0$ \cite{Schwinger:1951nm}.
 \item Parameter integral representations of $S^1$ are also known for specific one-dimensional field inhomogeneities, namely $E(t)\sim\cosh^{-2}(t/\tau)$ and $B({\rm x})\sim\cosh^{-2}({\rm x}/\lambda)$, with reference scales $\tau$ and $\lambda$, respectively \cite{Cangemi:1995ee}.
\end{itemize}
To perform the traces in Eq.~\eqref{eq:S1A} for a given field configuration $A_\text{cl.}^\mu(x)$, we need the eigenvalues of the exponentiated differential operators $D_{\text{cl.}}^2+\frac{e}{2}\sigma^{\mu\nu}F_{\text{cl.}\mu\nu}$ and $\partial^2$.

Subsequently, we stick to a {\it constant electromagnetic fields} and evaluate $S^1$ explicitly for this case.
For $F^{\mu\nu}_\text{cl.}=\text{const.}$, the expressions $D_{\text{cl.}}^2$ and $\frac{e}{2}\sigma^{\mu\nu}F_{\text{cl.}\mu\nu}$ in the argument of the first exponential in Eq.~\eqref{eq:S1A} of course commute, such that ${\rm Tr}\,{\rm e}^{{\rm i}(D_{\text{cl.}}^2+\frac{e}{2}\sigma^{\mu\nu}F_{\text{cl.}\mu\nu})T}={\rm Tr}\,{\rm e}^{{\rm i}\frac{e}{2}\sigma^{\mu\nu}F_{\text{cl.}\mu\nu}T}\,{\rm e}^{{\rm i}D_{\text{cl.}}^2T}$.
It can moreover be shown that in this case the entire background field dependence can be expressed in terms of the Lorentz and gauge invariant quantities ${\cal F}=\frac{1}{4}F^{\phantom{\mu}}_{\text{cl.}\mu\nu}F_\text{cl.}^{\mu\nu}=\frac{1}{2}(\vec{B}^2-\vec{E}^2)$ and ${\cal G}=\frac{1}{4}F^{\phantom{\mu}}_{\text{cl.}\mu\nu}{}^*F_\text{cl.}^{\mu\nu}=-\vec{B}\cdot\vec{E}$,
with dual field strength tensor $^*F^{\mu\nu}=\frac{1}{2}\epsilon^{\mu\nu\alpha\beta}F_{\alpha\beta}$; note that in our metric convention we have $F_{\text{cl.}0i}=-E_i$ and $F_{\text{cl.}ij}=\epsilon_{ijk}B_k$.
The reason for this is as follows:
Due to translational invariance in constant fields the effective action as well as the corresponding Lagrangian cannot explicitly depend on any space-time coordinate $x^\mu$ and derivative $\partial_\mu$.
Gauge invariance demands the field dependence to be entirely via $F_\text{cl.}^{\mu\nu}$. The fact that the effective Lagrangian is a scalar quantity and Lorentz invariance demand all Minkowski indices to be fully contracted.
Finally, all Lorentz-contracted monomials of the field strength tensor are reducible to $\cal F$ and $\cal G$ by means of the following identities \cite{Dittrich:2000zu}
\begin{align}
 F_\text{cl.}^{\mu\alpha}F_{\text{cl.}\alpha}^{\nu\phantom{\mu}} - {}^*F_\text{cl.}^{\mu\alpha}{}^*F_{\text{cl.}\alpha}^{\nu\phantom{\mu}} &= 2{\cal F}g^{\mu\nu} , \nonumber\\
 F_\text{cl.}^{\mu\alpha}{}^*F_{\text{cl.}\alpha}^{\nu\phantom{\mu}} = {}^*F_\text{cl.}^{\mu\alpha}F_{\text{cl.}\alpha}^{\nu\phantom{\mu}} &= {\cal G}g^{\mu\nu} . \label{eq:ids}
\end{align}
More precisely, as $S_\text{eff.}$ is a scalar quantity and $\cal G$ transforms as a pseudoscalar under Lorentz transformations, we actually have $S_\text{eff.}({\cal F},{\cal G}^2)$.

The trace over spinor indices is most conveniently performed by
noting that $(\frac{1}{2}\sigma^{\mu\nu}F_{\text{cl.}\mu\nu})^2=\frac{1}{8}\{\sigma^{\mu\nu},\sigma^{\alpha\beta}\}F_{\text{cl.}\mu\nu}F_{\text{cl.}\alpha\beta}=2({\cal F}+{\rm i}\gamma_5{\cal G})$ and the fact that the eigenvalues of $\gamma_5$ are doubly degenerate and given by $\pm1$ \cite{Schwinger:1951nm}. The latter property follows straightforwardly from $(\gamma_5)^2=\mathbb{1}$ and ${\rm tr}_\gamma\gamma_5=0$.
In turn, the eigenvalues of $(\frac{1}{2}\sigma^{\mu\nu}F_{\text{cl.}\mu\nu})^2$ are also doubly degenerate and read $2({\cal F}\pm{\rm i}{\cal G})$.
Taking into account that $\sigma^{\mu\nu}$ is traceless, it is then obvious that the four eigenvalues of $\frac{1}{2}\sigma^{\mu\nu}F_{\text{cl.}\mu\nu}$ are given by $\pm\sqrt{2({\cal F}\pm{\rm i}{\cal G})}$, and correspondingly
\begin{equation}
 {\rm tr}_\gamma\,{\rm e}^{{\rm i}\frac{e}{2}\sigma^{\mu\nu}F_{\text{cl.}\mu\nu}T}=2\,{\rm Re}\sum_{s=\pm1}{\rm e}^{-{\rm i}se\sqrt{2({\cal F}+{\rm i}{\cal G})}\,T}\,. \label{eq:trgamma}
\end{equation}
Therewith, $S^1$ can be expressed as
\begin{equation}
 S^1({\cal F},{\cal G}^2)=\frac{\rm i}{2}\int_{-{\rm i}/\Lambda^2}^\infty\frac{{\rm d}T}{T}\,{\rm e}^{-{\rm i}m^2T}\biggl\{{\rm tr}_x\,{\rm e}^{{\rm i}D_{\text{cl.}}^2T}\,2\,{\rm Re}\sum_{s=\pm1}{\rm e}^{-{\rm i}se\sqrt{2({\cal F}+{\rm i}{\cal G})}\,T}-4\,{\rm tr}_x\,{\rm e}^{{\rm i}\partial^2 T}\biggr\}\,. \label{eq:S1Aconst}
\end{equation}

To determine $S^1$ for generic constant electromagnetic fields explicitly, we make use of the following ``trick'':
We perform the calculation for the special case of $\vec{E}\parallel\vec{B}$, for which both $\cal F$ and $\cal G$ are nonzero and later express the result in terms of $\cal F$ and $\cal G$.
For the sake of definiteness we moreover assume $B>E$, such that there is a one-to-one correspondence between the quantities $E$, $B$ and $\cal F$, $\cal G$.
More specifically, we have
\begin{equation}
 E\leftrightarrow a:=\bigl(\sqrt{{\cal F}^2+{\cal G}^2}-{\cal F}\bigr)^{1/2} \quad\text{and} \quad B\leftrightarrow b:=\bigl(\sqrt{{\cal F}^2+{\cal G}^2}+{\cal F}\bigr)^{1/2} \,. \label{eq:ab}
\end{equation}
By means of Eq.~\eqref{eq:ab}, our result for $S^1$ written in terms of $E$ and $B$ can then be straightforwardly generalized to arbitrary oriented constant electric and magnetic fields of arbitrary strength.

Let us now turn to the explicit calculation.
Without loss of generality we choose $A_\text{cl.}^\mu(x)=(0,0,B{\rm x},-Et)$, so that we have $\vec{B}=B\vec{e}_{\rm z}$, $\vec{E}=E\vec{e}_{\rm z}$ and $D_\text{cl.}^2=\partial_{\rm x}^2+(\partial_{\rm y}-{\rm i}eB{\rm x})^2-\partial_t^2+(\partial_{\rm z}+{\rm i}eEt)^2$.
The second trace in Eq.~\eqref{eq:S1Aconst} can be evaluated straightforwardly in momentum space as $\partial_\mu={\rm i}\hat p_\mu$,
with momentum operator $\hat p_\mu$ fulfilling $\hat p_\mu|p_\mu\rangle=p_\mu|p_\mu\rangle$ (no summation over repeated indices). Note that $\langle p'|p\rangle=(2\pi)^4\delta^{(4)}(p'-p)$ and $\langle x'|x\rangle=\delta^{(4)}(x'-x)$.
With $|p\rangle\equiv|p_0\rangle|p_1\rangle|p_2\rangle|p_3\rangle$ we obtain
\begin{equation}
 {\rm tr}_x\,{\rm e}^{{\rm i}\partial^2T}=\int\frac{{\rm d}^4p}{(2\pi)^4}\langle p|{\rm e}^{{\rm i}\partial^2T}|p\rangle
 =L^4\int\frac{{\rm d}^4p}{(2\pi)^4}{\rm e}^{-{\rm i}p^2 T}=-{\rm i}\,\frac{L^4}{16\pi^2}\frac{1}{T^2}\,, \label{eq:TrFree}
\end{equation}
where we employed the identity $\langle p_\mu|p_\mu\rangle=L$ (no summation over repeated indices), with space-time extension $L$.
To perform the first trace in Eq.~\eqref{eq:S1Aconst}, we employ a mixed momentum and position space representation, i.e., 
${\rm tr}_x\{\ldots\}=\int{\rm d}t\int{\rm dx}\int\frac{{\rm d}p_{\rm y}}{2\pi}\int\frac{{\rm d}p_{\rm z}}{2\pi}
\langle t|\langle{\rm x}|\langle p_{\rm y}|\langle p_{\rm z}|\ldots|p_{\rm z}\rangle|p_{\rm y}\rangle|{\rm x}\rangle|t\rangle$.
In turn, we have
\begin{align}
{\rm tr}_x\,{\rm e}^{{\rm i}D_{\text{cl.}}^2T}
&=L\int\frac{{\rm d}p_{\rm z}}{2\pi}\int{\rm d}t\,\langle t|{\rm e}^{{\rm i}eE[\frac{1}{eE}(-{\rm i}\partial_t)^2-eE(t+\frac{p_{\rm z}}{eE})^2]T}|t\rangle \nonumber\\
&\hspace*{2cm}\times L\int\frac{{\rm d}p_{\rm y}}{2\pi}\int{\rm dx}\,\langle{\rm x}|\,{\rm e}^{-{\rm i}eB[\frac{1}{eB}(-{\rm i}\partial_{\rm x})^2+eB({\rm x}-\frac{p_{\rm y}}{eB})^2]T}|{\rm x}\rangle \,,
\end{align}
{and after shifting and rescaling $t\to\tilde t=\sqrt{eE}(t+\frac{p_{\rm z}}{eE})$ and ${\rm x}\to\tilde {\rm x}=\sqrt{eB}({\rm x}-\frac{p_{\rm y}}{eB})$,
\begin{align}
{\rm tr}_x\,{\rm e}^{{\rm i}D_{\text{cl.}}^2T}
&=\frac{LP_{\rm z}}{2\pi}\int{\rm d}\tilde t\,\langle\tilde t|{\rm e}^{{\rm i}2eE[\frac{1}{2}(-{\rm i}\partial_{\tilde t})^2+\frac{1}{2}{\rm i}^2\tilde t^2]T}|\tilde t\rangle \nonumber\\
&\hspace*{2cm}\times
\frac{LP_{\rm y}}{2\pi}\int{\rm d}\tilde{\rm x}\,\langle\tilde{\rm x}|\,{\rm e}^{-{\rm i}2eB[\frac{1}{2}(-{\rm i}\partial_{\tilde{\rm x}})^2+\frac{1}{2}\tilde{\rm x}^2]T}|\tilde{\rm x}\rangle \,,
\label{eq:trD}
\end{align}
where we employed $\int{\rm d}p_i=\lim_{P_i\to\infty} P_i$, and the limit $\lim_{P_i\to\infty}$ is implicitly understood.

It is now helpful to note that the expressions in the squared brackets in Eq.~\eqref{eq:trD} amount to the (position space) Hamilton operator of a quantum harmonic oscillator in one dimensions,
$\hat H=\frac{1}{2m}\hat p_{\rm x}^2+\frac{1}{2}m\omega^2\hat{\rm x}^2$, whose eigenvalues are given by $E_n=\omega(n+\frac{1}{2})$, with $n\in\mathbb{N}_0$.
For the expression in the lower line of Eq.~\eqref{eq:trD} we have $\omega=1$.
The expression in the upper line of Eq.~\eqref{eq:trD} is to be understood as harmonic oscillator with its frequency analytically continued to $\omega={\rm i}$. This becomes obvious from the fact that as a direct consequence of $S^1=S^1({\cal F},{\cal G}^2)$, the results for the effective action in a purely magnetic field and a purely electric field are related by an analytical continuation from $B\leftrightarrow-{\rm i}E$ \cite{Jentschura:2001qr} (cf. also below).
Note that seemingly the choice of $\omega=-{\rm i}$ would be equally justified.
However, only $\omega={\rm i}$ results in a convergent result, and thus is compatible with the limit of $eE\to0$ in the first line of Eq.~\eqref{eq:->zerofield} below.
Correspondingly, we have
\begin{equation}
{\rm tr}_x\,{\rm e}^{{\rm i}D_{\text{cl.}}^2T}
=\frac{LP_{\rm z}}{2\pi}\sum_{n'=0}^\infty \rho'(n')\,{\rm e}^{-eE(2n'+1)T}\,
\frac{LP_{\rm y}}{2\pi}\sum_{n=0}^\infty \rho(n)\,{\rm e}^{-{\rm i}eB(2n+1)T} , \label{eq:TrB}
\end{equation}
where $\rho'(n')$ and $\rho(n)$ denote the densities of states of the discrete eigenvalues.
They can be determined by considering the limits of $eE\to0$ and $eB\to0$, respectively,
\begin{align}
 \lim_{eE\to0}\frac{LP_{\rm z}}{2\pi}\sum_{n'=0}^\infty \rho'(n')\,{\rm e}^{-eE(2n'+1)T}&=L^2\int\frac{{\rm d}p_{\rm z}}{2\pi}\int\frac{{\rm d}p_t}{2\pi}\,{\rm e}^{-(p_{\rm z}^2+p_t^2)T} \,, \nonumber\\
 \lim_{eB\to0}\frac{LP_{\rm y}}{2\pi}\sum_{n=0}^\infty \rho(n)\,{\rm e}^{-{\rm i}eB(2n+1)T}&=L^2\int\frac{{\rm d}p_{\rm x}}{2\pi}\int\frac{{\rm d}p_{\rm y}}{2\pi}\,{\rm e}^{-{\rm i}(p_{\rm x}^2+p_{\rm y}^2)T} \,. \label{eq:->zerofield}
\end{align}
where we made use of the identity $\int\frac{{\rm d}p_{\rm z}}{2\pi}\int\frac{{\rm d}p_t}{2\pi}\,{\rm e}^{-{\rm i}(p_{\rm z}^2-p_t^2)T}=\int\frac{{\rm d}p_{\rm z}}{2\pi}\int\frac{{\rm d}p_t}{2\pi}\,{\rm e}^{-(p_{\rm z}^2+p_t^2)T}$.
The latter line can be expressed as
\begin{equation}
 \lim_{eB\to0} \sum_{n=0}^\infty \Delta p_\perp^2 \frac{\rho(n)}{2eB}\,{\rm e}^{-{\rm i}n\Delta p_\perp^2T} =\frac{L}{P_{\rm y}}\frac{1}{2}\int_0^\infty{\rm d}p_\perp^2 {\rm e}^{-{\rm i}p_\perp^2T} ,
\end{equation}
where we defined $p_\perp^2\equiv p_{\rm x}^2+p_{\rm y}^2$ and $\Delta p_\perp^2=2eB$. Herefrom we infer $\rho(n)=eB\frac{L}{P_{\rm y}}$.  An analogous manipulation of the expression in the first line yields $\rho'(n')=eE\frac{L}{P_{\rm z}}$.
In turn, Eq.~\eqref{eq:TrB} becomes
\begin{align}
{\rm tr}_x\,{\rm e}^{{\rm i}D_{\text{cl.}}^2T}
&=\frac{L^4}{4\pi^2}\sum_{n'=0}^\infty eE\,{\rm e}^{-eE(2n'+1)T}\sum_{n=0}^\infty eB\,{\rm e}^{-{\rm i}eB(2n+1)T} \nonumber\\
&=-{\rm i}\,\frac{L^4}{16\pi^2}\frac{1}{T^2}\frac{(eET)(eBT)}{\sinh(eET)\sin(eBT)}\,. \label{eq:TrB2}
\end{align}
This expression is clearly invariant under the transformation $B\leftrightarrow-{\rm i}E$.
Noting that $2({\cal F}+{\rm i}{\cal G})=(B-{\rm i}E)^2$, it is moreover obvious that Eq.~\eqref{eq:trgamma} can be written as
\begin{equation}
 {\rm tr}_\gamma\,{\rm e}^{{\rm i}\frac{e}{2}\sigma^{\mu\nu}F_{\text{cl.}\mu\nu}T}=2\,{\rm Re}\sum_{s=\pm1}{\rm e}^{-{\rm i}se(B-{\rm i}E)T}=4\cosh(eET)\cos(eBT)\,. 
\end{equation}
Putting everything together, we finally obtain
\begin{align}
 S^1({\cal F},{\cal G}^2)&=\frac{L^4}{8\pi^2}\int_{-{\rm i}/\Lambda^2}^\infty\frac{{\rm d}T}{T^3}\,{\rm e}^{-{\rm i}m^2T}\biggl\{\frac{(eaT)(ebT)}{\tanh(eaT)\tan(ebT)}-1\biggr\} \nonumber\\
 &=-\frac{L^4}{8\pi^2}\int_{1/\Lambda^2}^{\infty}\frac{{\rm d}T}{T^3}\,{\rm e}^{-m^2T}\biggl\{\frac{(eaT)(ebT)}{\tan(eaT)\tanh(ebT)}-1\biggr\}\,. \label{eq:S1Ab}
\end{align}
where we employed a contour rotation ${\rm i}T\to T$ in the last step.
Equation~\eqref{eq:S1Ab} provides us with a compact expression of the unrenormalized on-loop effective action in constant electromagnetic fields.
The corresponding Lagrangian density is ${\cal L}^1({\cal F},{\cal G}^2)=S^1({\cal F},{\cal G}^2)/L ^4$.
Note that
\begin{equation}
 \frac{(eaT)(ebT)}{\tan(eaT)\tanh(ebT)}-1=\frac{2}{3}{\cal F}(eT)^2-\Bigl(\frac{4}{45}{\cal F}^2+\frac{7}{45}{\cal G}^2\Bigr)(eT)^4+{\cal O}(T^6)\,, \label{eq:exp}
\end{equation}
which implies that Eq.~\eqref{eq:S1Ab} contains a log-type divergence for $\Lambda\to\infty$ in the term $\sim {\cal F}$. All other contributions are finite in this limit.
However, let us emphasize once again that so far we worked with unrenomalized fields and not at all bothered about renormalization. 

In a next step we want go over to {\it renormalized quantities}. To this end we isolate this term. Upon its subtraction, the expression in Eq.~\eqref{eq:S1Ab} is finite for $\Lambda\to\infty$ and we can explicitly take the limit and set the lower integration bound to zero. This results in
\begin{align}
 {\cal L}^1({\cal F},{\cal G}^2)=&-{\cal F}\frac{\alpha}{3\pi}\int_{1/\Lambda^2}^\infty\frac{{\rm d}T}{T}\,{\rm e}^{-m^2T}\nonumber\\
 &-\frac{1}{8\pi^2}\int_{1/\Lambda^2}^{\infty}\frac{{\rm d}T}{T^3}\,{\rm e}^{-m^2T}\biggl\{\frac{(eaT)(ebT)}{\tan(eaT)\tanh(ebT)}-\frac{2}{3}(eT)^2{\cal F}-1\biggr\} \,, \label{eq:S1Ac}
\end{align}
with $\alpha=\frac{e^2}{4\pi}$ and $\int_{1/\Lambda^2}^\infty\frac{{\rm d}T}{T}\,{\rm e}^{-m^2T}=\ln(\frac{\Lambda^2}{m^2})-\gamma_E+{\cal O}(\frac{m^2}{\Lambda^2})$,
where $\gamma_E$ denotes the Euler-Mascheroni constant.
Together with the Maxwell term ${\cal L}_\text{MW}=-{\cal F}$ the effective Lagrangian can be expressed as
\begin{align}
 {\cal L}_\text{eff.}({\cal F},{\cal G}^2)=&-{\cal F}\biggl\{1+\frac{\alpha}{3\pi}\Bigl[\ln\Bigl(\frac{\Lambda^2}{m^2}\Bigr)-\gamma_E\Bigr]\biggr\} \nonumber\\
 &-\frac{1}{8\pi^2}\int_{0}^{\infty}\frac{{\rm d}T}{T^3}\,{\rm e}^{-m^2T}\biggl\{\frac{(eaT)(ebT)}{\tan(eaT)\tanh(ebT)}-\frac{2}{3}(eT)^2{\cal F}-1\biggr\}\,. \label{eq:Seff1}
\end{align}
Let us now impose the renormalization condition that the term $\sim{\cal F}$ in Eq.~\eqref{eq:Seff1} matches the physical (measured) Maxwell term.
For this purpose we introduce a wave function renormalization factor
\begin{equation}
 Z^{-1}\equiv 1+\frac{\alpha}{3\pi}\Bigl[\ln\Bigl(\frac{\Lambda^2}{m^2}\Bigr)-\gamma_E\Bigr]\,,
\end{equation}
and define the renormalized field strength tensor by $F_\text{cl.,R}^{\mu\nu}=Z^{-1/2}F_\text{cl.}^{\mu\nu}$, such that the renormalized four-potential is given by $A_\text{cl.,R}^\mu=Z^{-1/2}A_\text{cl.}^\mu$.
For the interaction vertex in the Lagrangian~\eqref{eq:L_nQED} this rescaling implies $eA_{\text{cl.}}^\mu\bar\psi \gamma_\mu\psi\to Z^{1/2}eA_{\text{cl.,R}}^\mu\bar\psi \gamma_\mu\psi$.
Demanding it to be given by $e_\text{R} A_{\text{cl.,R}}^\mu\bar\psi \gamma_\mu\psi$, i.e., to match the physical interaction, we infer $e_\text{R}=Z^{1/2}e$.
Correspondingly, we have
\begin{gather}
  F^{\mu\nu}_\text{R}=Z^{-1/2}F^{\mu\nu}=F^{\mu\nu}_\text{R}(m)\,, \nonumber\\
  e_\text{R}=Z^{1/2}e=e_\text{R}(m)\,,
\end{gather}
but $e_\text{R}F^{\mu\nu}_\text{R}=eF^{\mu\nu}$ independent of the renormalization scale $\mu=m$. Note that the choice of $\mu=m$ is conventionally referred to as ``on-shell'' renormalization.
For this choice we have $\frac{e_\text{R}^2(m)}{4\pi}=\alpha_\text{R}(m)\simeq\frac{1}{137}$.
In turn, the (on-shell) renormalized effective Lagrangian reads
\begin{align}
 {\cal L}^\text{R}_\text{eff.}({\cal F}_\text{R},{\cal G}_\text{R}^2)=&-{\cal F}_R \nonumber\\
 &-\frac{1}{8\pi^2}\int_{0}^{\infty}\frac{{\rm d}T}{T^3}\,{\rm e}^{-m^2T}\biggl\{\frac{(e_\text{R}a_\text{R}T)(e_\text{R}b_\text{R}T)}{\tan(e_\text{R}a_\text{R}T)\tanh(e_\text{R}b_\text{R}T)}-\frac{2}{3}(e_\text{R}T)^2{\cal F}_\text{R}-1\biggr\}\,. \label{eq:L_HE}
\end{align}

The expression in Eq.~\eqref{eq:L_HE} amounts to the renowned {\it Heisenberg-Euler effective Lagrangian} \cite{Heisenberg:1935qt,Schwinger:1951nm} first derived by Werner Heisenberg and Hans Euler and published eighty years ago in 1936 \cite{Heisenberg:1935qt}. The analogous result for scalar QED can be obtained along the same lines and was first derived by Victor Weisskopf \cite{Weisskopf} and also published in 1936; see \cite{Dunne:2004nc} for a review.
All our subsequent considerations will be based upon Eq.~\eqref{eq:L_HE}, i.e., we will exclusively work with renormalized quantities. However, in order to keep notations compact we will suppress the label ``$\rm R$'' in the following.

Let us emphasize once again that the one-loop effective Lagrangian~\eqref{eq:L_HE} accounts for all orders in ${\cal F}$ and ${\cal G}^2$, and thus is fully nonperturbative in the combined parameter $eF_\text{cl.}^{\mu\nu}$.
Due to ``Furry's theorem'' \cite{Furry:1937qvr} (charge conjugation invariance of QED), the effective Lagrangian is moreover even in the elementary charge $e$, and in turn even in $eF_\text{cl.}^{\mu\nu}$.
To lowest order in a perturbative expansion of Eq.~\eqref{eq:L_HE} -- counting ${\cal O}({\cal F})={\cal O}({\cal G})={\cal O}(F^2)$ -- we obtain [cf. Eq.~\eqref{eq:exp}]
\begin{align}
 {\cal L}_\text{eff.}({\cal F},{\cal G}^2)&=-{\cal F} + \frac{1}{8\pi^2}\int_{0}^{\infty}\frac{{\rm d}T}{T^3}\,{\rm e}^{-m^2T}\biggl[\frac{4}{45}{\cal F}(eT)^4+\frac{7}{45}{\cal G}^2(eT)^4+ {\cal O}\bigl((eFT)^6\bigr)\biggr]  \nonumber\\
 &=-{\cal F} + \underbrace{\frac{8}{45}\frac{\alpha^2}{m^4}}_{=: c_1}{\cal F}^2+\underbrace{\frac{14}{45}\frac{\alpha^2}{m^4}}_{=: c_2}{\cal G}^2+ m^4{\cal O}\Bigl(\bigl(\tfrac{\alpha F^2}{m^4}\bigr)^3\Bigr) \,. \label{eq:L_HEpert} 
\end{align}
\vspace*{-0.3cm}\\
\hspace*{0.37cm}\centerline{\includegraphics[height=2.1cm]{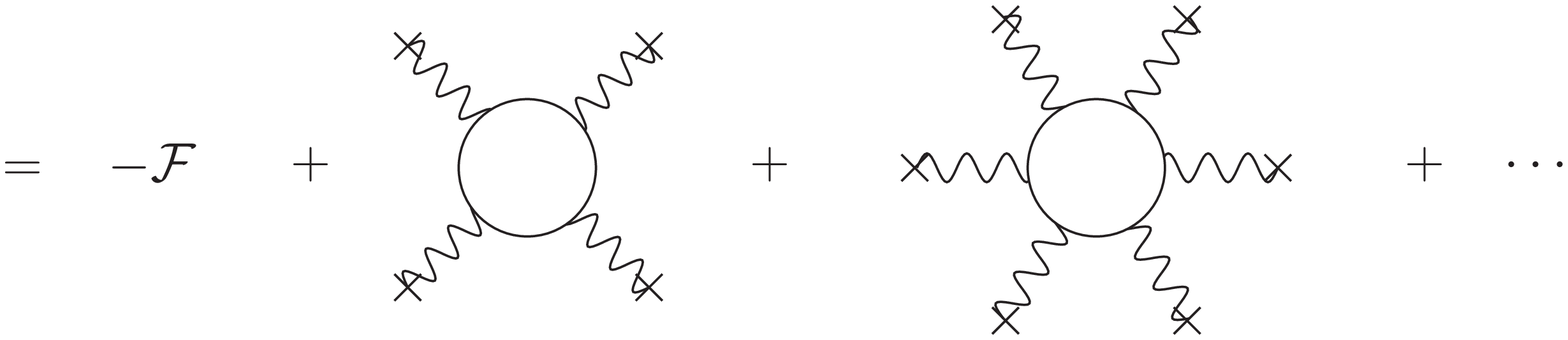}}\vspace{2mm}
The expression $c_1{\cal F}^2+c_2{\cal G}^2\sim F^4$ obviously corresponds to the first loop diagram exhibiting four couplings to the electromagnetic field. 
For completeness, note that by means of an expansion of the integrand of the proper time integral the explicit expression of any perturbative one-loop interaction among constant electromagnetic fields can be obtained; cf. also \cite{Dunne:2004nc}.

Even though the above results are strictly speaking only valid for constant electromagnetic fields, they can also be adopted for {\it slowly varying fields}:
\newcommand{\lambdabar}{{\mkern0.75mu\mathchar '26\mkern -9.75mu\lambda}}
Vacuum fluctuations of virtual electrons and positrons typically probe distances of the order of the (reduced) Compton wavelength of the electron $\lambdabar_c=\frac{1}{m}\approx3.9\cdot10^{-13}{\rm m}$; the associated time scale is the Compton time $\tau_c\approx 1.3\cdot 10^{-21}{\rm s}$.
In turn, $\lambdabar_c$ provides us with a natural reference scale to classify spatial variations.
Consider now a derivative expansion around the constant field result. As the derivatives are rendered dimensionless by $\lambdabar_c$, terms involving $n\in\mathbb{N}_0$ derivatives scale as $(\lambdabar_c \partial_{\rm x})^n\sim(\frac{\omega}{m})^n$, where $\omega$ is the typical momentum scale of variation of the field configuration under consideration.
Hence, for {\it slowly varying fields}, characterized by $\frac{\omega}{m}\ll1$, all derivative terms can be neglected and we can employ the substitution $F^{\mu\nu}\to F^{\mu\nu}(x)$ in the constant field result for ${\cal L}_\text{eff.}$ in Eqs.~\eqref{eq:L_HE} and \eqref{eq:L_HEpert}. This substitution amounts to a {\it locally constant field approximation} (LCFA).

Let us emphasize that in fact most of the field configurations attainable in the laboratory vary on scales much larger than $\lambdabar_c$ and thus fall into this category.
Correspondingly, the LCFA allows for reliable analytical insights into a variety of experimental relevant field configurations, without having to determine the effective action in the dedicated inhomogeneous electromagnetic field profile.

\section{Vacuum birefringence as an optical signature of quantum vacuum nonlinearity}

In the following, we adopt the LCFA and study {\it light propagation} in the quantum vacuum subjected to classical electromagnetic fields.
As ${\cal L}_\text{eff.}={\cal L}_\text{eff.}(F^{\mu\nu}_\text{cl.})=-{\cal F}+{\cal L}^1({\cal F},{\cal G}^2)$, the equations of motion~\eqref{eq:EulerLagrange} for the electromagnetic field in the quantum vacuum are given by 
\begin{equation}
 \partial_\mu\frac{{\cal L}_\text{eff.}}{\partial F_{\text{cl.}\mu\nu}}=0 \quad \leftrightarrow \quad \partial_\mu\biggl(F_\text{cl.}^{\mu\nu}-\frac{\partial{\cal L}^1}{\partial {\cal F}}\,F_\text{cl.}^{\mu\nu}-\frac{\partial{\cal L}^1}{\partial {\cal G}}\,{}^*F_\text{cl.}^{\mu\nu}\biggr)=0 \,. \label{eq:EoM0}
\end{equation}
Equation~\eqref{eq:EoM0} constitutes the ``true'' {\it quantum Maxwell equations} (at one-loop order) amounting to the {\it true} equations of motion governing the dynamics of classical electromagnetic field in the quantum vacuum far away from any classical sources.
Note that these equations are no longer linear in $F^{\mu\nu}_\text{cl.}$ and thus -- at least in principle -- violate the superposition principle of classical electrodynamics; cf. below Eq.~\eqref{eq:EulerLagrange} above.
Due to the fact that these nonlinearities are suppressed parametrically by powers of $\frac{eE}{m^2}=\frac{E[{\rm V/m}]}{1.3\cdot10^{18}}$ and $\frac{eB}{m^2}=\frac{B[{\rm T}]}{4\cdot10^9}$, and all the macroscopic electromagnetic fields available in the laboratory clearly fulfill $\{\frac{eE}{m^2},\frac{eB}{m^2}\}\ll1$, these nonlinearities are typically extremely tiny, preserving the applicability of the Maxwell's classical equations for essentially all practical purposes.
As will be shown explicitly below they can give rise to striking consequences for photon propagation in electromagnetic fields. 
From Eq.~\eqref{eq:L_HEpert} we infer $\frac{\partial{\cal L}^1}{\partial {\cal F}}=2c_1{\cal F}\bigl[1+{\cal O}\bigl((\frac{eF}{m^2})^2\bigr)\bigr]$ and $\frac{\partial{\cal L}^1}{\partial {\cal G}}=2c_2{\cal G}\bigl[1+{\cal O}\bigl((\frac{eF}{m^2})^2\bigr)\bigr]$, such that to cubic order in the field strength we have
\begin{equation}
 \partial_\mu\bigl[(1-2c_1{\cal F})F_\text{cl.}^{\mu\nu}-2c_2{\cal G}{}^*F_\text{cl.}^{\mu\nu}\bigr]=0\,. \label{eq:EoM}
\end{equation}
A generic field strength tensor $F^{\mu\nu}$ fulfills the Bianchi identity $\partial^\alpha F^{\beta\gamma}+\partial^\beta F^{\gamma\alpha}+\partial^\gamma F^{\alpha\beta}=0$
$\leftrightarrow$ $\partial_\mu{}^*F^{\mu\nu}=0$. With its help Eq.~\eqref{eq:EoM} can be written as
\begin{equation}
 \partial_\mu\bigl[(1-2c_1{\cal F})F_\text{cl.}^{\mu\nu}\bigr]
 -2c_2\bigl(\partial_\mu{\cal G}\bigr){}^*F_\text{cl.}^{\mu\nu}=0\,. \label{eq:EoM1}
\end{equation}

In a next step we decompose the field strength tensor as $F_\text{cl.}^{\mu\nu}\to F_\text{cl.}^{\mu\nu}+f^{\mu\nu}(x)$ into a constant background field $F_\text{cl.}^{\mu\nu}$ and a probe photon field $f^{\mu\nu}(x)$,
and linearize the equations of motion~\eqref{eq:EoM} in $f^{\mu\nu}$ \cite{BialynickaBirula:1970vy,Dittrich:2000zu}, i.e., neglect self-interactions of the probe photon field.
Note that under the above decomposition we have
\begin{align}
 {\cal F}&\to{\cal F}+\frac{1}{2}F_{\text{cl.}\mu\nu}f^{\mu\nu}(x)+{\cal O}(f^2) \,, \nonumber\\
 {\cal G}&\to{\cal G}+\frac{1}{2}{}^*F_{\text{cl.}\mu\nu}f^{\mu\nu}(x)+{\cal O}(f^2) \,.
\end{align}
Correspondingly, we obtain the following linearized equations of motion
\begin{equation}
 \bigl(1-2c_1{\cal F}\bigr)\partial_\mu f^{\mu\nu}(x)
 -c_1 F_{\text{cl.}\alpha\beta}F_\text{cl.}^{\mu\nu}\partial_\mu f^{\alpha\beta}(x)
 -c_2{}^*F_{\text{cl.}\alpha\beta}{}^*F_\text{cl.}^{\mu\nu}\partial_\mu f^{\alpha\beta}(x)=0 \,.  \label{eq:EoM2}
\end{equation}
In order to solve them, we first go to momentum space, i.e., $a^\mu(x)=\int\frac{{\rm d}^4k}{(2\pi)^4}{\rm e}^{{\rm i}kx}a^\mu(k)$ $\leftrightarrow$
$f^{\mu\nu}(x)={\rm i}\int\frac{{\rm d}^4k}{(2\pi)^4}{\rm e}^{{\rm i}kx}[k^\mu a^\nu(k)-k^\nu a^\mu(k)]$, and work in Lorenz gauge, $\partial_\mu a^\mu(x)=0$ $\leftrightarrow$
$k_\mu a^\mu(k)=0$.
In momentum space the equations of motion can be cast into the following form,
\begin{equation}
 \bigl(1-2c_1{\cal F}\bigr)k^2 a^{\nu}(k)
 -2c_1 F_{\text{cl.}\alpha\beta}F_\text{cl.}^{\mu\nu}k_\mu k^\alpha a^\beta(k)
 -2c_2{}^*F_{\text{cl.}\alpha\beta}{}^*F_\text{cl.}^{\mu\nu}k_\mu k^\alpha a^\beta(k)=0 \,.  \label{eq:EoM3}
\end{equation}
Let us now search for nontrivial solutions of this equation.
Our strategy is to use the following ans\"atze
\begin{align}
 a_1^\mu(k)&\sim F_\text{cl.}^{\mu\nu}k_\nu\equiv(F_\text{cl.}k)^\mu=(\vec{k}\cdot\vec{E},\vec{k}\times\vec{B}+\omega\vec{E}) \,, \nonumber\\
 a_2^\mu(k)&\sim{}^*F_\text{cl.}^{\mu\nu}k_\nu\equiv({}^*F_\text{cl.}k)^\mu=(\vec{k}\cdot\vec{B},-\vec{k}\times\vec{E}+\omega\vec{B}) \,, \label{eq:as}
\end{align}
with $k^\mu=(\omega,\vec{k})$ and the contraction identities~\eqref{eq:ids}.
These ans\"atze obviously fulfill $k_\mu a_i^\mu(k)=0$, i.e., are compatible with the Lorenz gauge condition.
Also note that $(F_\text{cl.}k)^2=\omega^2\vec{E}^2+\vec{k}^2\vec{B}^2-(\vec{k}\cdot\vec{E})^2-(\vec{k}\cdot\vec{B})^2+2\omega(\vec{k}\times\vec{B})\cdot\vec{E}$.
From Eq.~\eqref{eq:EoM3} it is obvious that the dispersion relations for the $a_i^\mu(k)$ are of the form $k^2=0+\alpha\,{\cal O}\bigl((\tfrac{eF}{m^2})^2\bigr)$.

Let us have a closer look on the physical meaning of the above ans\"atze. Our main focus is on their polarizations, defined by the directions of the associated electric field vectors. 
With the classical free Maxwell equations in the vacuum, $\partial_\mu F_\text{cl.}^{\mu\nu}=0$, we obtain
\begin{align}
 \text{position\ space:}\quad
 \begin{cases}
  \vec{e}_i(x)=-\vec\nabla a_i^0(x) -\partial_t\vec{a}_i(x) \\
  \vec{b}_i(x)=\vec\nabla\times\vec{a}_i(x)
 \end{cases}\hspace*{-0.2cm}, \nonumber\\
 \to\quad \text{momentum\ space:}\quad
 \begin{cases}
  \vec{e}_i(k)\sim-\vec{k} a_i^0(k) +\omega\vec{a}_i(k) \\
  \vec{b}_i(k)\sim\vec{k}\times\vec{a}_i(k)
\end{cases}\hspace*{-0.2cm}. \hspace*{0.2cm}
 \end{align}
The use of the classical Maxwell equations neglecting quantum corrections is justified here, as the quantum corrections only give rise to subleading corrections.
The electric $\vec{e}_i$ and magnetic $\vec{b}_i$ fields associated with the $a_i^\mu(k)$ in Eq.~\eqref{eq:as} read
\begin{align}
a_1^\mu(k):\quad
\begin{cases}
 \vec{e}_1&\sim-\vec{k}(\vec{k}\cdot\vec{E})+\omega(\vec{k}\times\vec{B})+\omega^2\vec{E} \\
 \vec{b}_1&\sim \omega(\vec{k}\times\vec{E})-\vec{k}^2\vec{B}+\vec{k}(\vec{k}\cdot\vec{B})
\end{cases}, \\
a_2^\mu(k):\quad
\begin{cases}
 \vec{e}_2&\sim-\vec{k}(\vec{k}\cdot\vec{B})-\omega(\vec{k}\times\vec{E})+\omega^2\vec{B} \\
 \vec{b}_2&\sim \omega(\vec{k}\times\vec{B})+\vec{k}^2\vec{E}-\vec{k}(\vec{k}\cdot\vec{E})
\end{cases}.
\end{align}
Note that $\vec{k}\cdot\vec{e}_1\sim\vec{k}\cdot\vec{e}_2\sim{\cal G}k^2$ and $\vec{e}_1\cdot\vec{e}_2\sim k^2\bigl[(\vec{k}\cdot\vec{E})(\vec{k}\cdot\vec{B})+\omega^2{\cal G}\bigr]$. This implies that for probe photons propagating on the light cone, i.e., fulfilling $k^2=0$, the $a_i^\mu(k)$ span the two transverse photon polarization modes. More specifically, they correspond to the two orthogonal linear polarization modes $\sim\vec{e}_i$ in a particular basis set by the background field configuration and the photons' wave vector. Of course, for the case of $k^2=0+\alpha\,{\cal O}\bigl((\tfrac{eF}{m^2})^2\bigr)$ this is still well-justified.

With the shorthand notations introduced in Eq.~\eqref{eq:as}, we can express Eq.~\eqref{eq:EoM3} as
\begin{equation}
 \bigl(1-2c_1{\cal F}\bigr)k^2 a^{\nu}(k)
 +2c_1(F_\text{cl.}k)^\nu k^\alpha F_{\text{cl.}\alpha\beta}  a^\beta(k)
 +2c_2({}^*F_\text{cl.}k)^\nu k^\alpha{}^*F_{\text{cl.}\alpha\beta} a^\beta(k)=0 \,.  \label{eq:EoM4}
\end{equation}
In a next step we plug the above ans\"atze into this equation.
After some rearrangements and using the contraction identities~\eqref{eq:ids}, the respective equations become
\begin{align}
 a_1^\mu(k):&\quad\bigl[\bigl(1-2c_1{\cal F}\bigr)k^2-2c_1(F_\text{cl.} k)^2\bigr](F_\text{cl.}k)^\nu+2c_2{\cal G}k^2({}^*F_\text{cl.}k)^\nu =0 \,, \nonumber\\
 a_2^\mu(k):&\quad\bigl[\bigl(1-2c_1{\cal F}+2c_2{\cal F}\bigr)k^2-2c_2(F_\text{cl.} k)^2\bigr]({}^*F_\text{cl.}k)^\nu-2c_1{\cal G}k^2(F_\text{cl.}k)^\nu =0 \,.
\end{align}
These equations can be solved with the ansatz $k^2=0+\Delta$, where $\Delta=\alpha\,{\cal O}\bigl((\tfrac{eF}{m^2})^2\bigr)$ (cf. also above).
Note that this solution procedure can systematically (recursively) be extended to higher orders.
We denote the solution of the equation for $a_i^\mu(k)$ by $\Delta_i$.
In fact, it is very easy to see that
\begin{equation}
 \Delta_i=2c_i(F_\text{cl.}k)^2+\alpha\,{\cal O}\bigl((\tfrac{eF}{m^2})^4\bigr)\,.
\end{equation}
Hence, the dispersion relations for the modes $a_i^\mu(k)$ are given by
\begin{equation}
 k^2=2c_i(F_\text{cl.}k)^2 +\alpha\,{\cal O}\bigl((\tfrac{eF}{m^2})^4\bigr) \quad\leftrightarrow\quad \omega\bigl(|\vec{k}|\bigr)=|\vec{k}|\bigl[1-c_i(F_\text{cl.}k)^2/\vec{k}^2\bigr] +\alpha\,{\cal O}\bigl((\tfrac{eF}{m^2})^4\bigr)   \,. \label{eq:disps}
\end{equation}
Taking into account $(F_\text{cl.}k)^2=(F_\text{cl.}k)^2\big|_{k^2=0}+(eF)^2\,{\cal O}\bigl((\tfrac{eF}{m^2})^4\bigr)$,
with $(F_\text{cl.}k)^2\big|_{k^2=0}=(\vec{k}\times\vec{E})^2+(\vec{k}\times\vec{B})^2-2|\vec{k}|\,\vec{k}\cdot(\vec{E}\times\vec{B})$ (cf. above), the dispersion relations~\eqref{eq:disps} can finally be written as
\begin{equation}
 \omega\bigl(|\vec{k}|\bigr)=|\vec{k}|\Bigl\{1-c_i\Bigl[E^2\sin^2\sphericalangle(\vec{k},\vec{E})+B^2\sin^2\sphericalangle(\vec{k},\vec{B})-2EB\,\hat{\vec{k}}\cdot\hat{\vec{s}}\,\Bigr]\Bigr\} +\alpha\,{\cal O}\bigl((\tfrac{eF}{m^2})^4\bigr)\,,
 \label{eq:disps1}
\end{equation}
with $\hat{\vec{k}}=\vec{k}/|\vec{k}|$, $E=|\vec{E}|$, $B=|\vec{B}|$, and unit Poynting vector $\hat{\vec{s}}=\frac{\vec{E}\times\vec{B}}{EB}$.

In turn, the probe photons' group and phase velocities derived from Eq.~\eqref{eq:disps1} agree with each other, i.e., $v_{\text{gr},i}=\frac{{\rm d}\omega}{{\rm d}|\vec{k}|}=v_{\text{ph},i}=\frac{\omega}{|\vec{k}|}=:v_i$, and are given by
\begin{align}
 \left\{\!\!\begin{array}{c}
  v_1 \\ v_2
 \end{array}\!\!\right\}
 =1-\frac{\alpha}{\pi}\frac{1}{90}\!
  \left\{\!\!\begin{array}{c}
  7 \\ 4
 \end{array}\!\!\right\}\!
 \biggl[\Bigl(\frac{eE}{m^2}\Bigr)^2\sin^2\sphericalangle(\vec{k},\vec{E})+\Bigl(\frac{eB}{m^2}\Bigr)^2\sin^2\sphericalangle(\vec{k},\vec{B})-2\frac{eE}{m^2}\frac{eB}{m^2}\,\hat{\vec{k}}\cdot\hat{\vec{s}}\,\biggr]\,,
 \label{eq:vis}
\end{align}
where the neglected higher order corrections are $\sim\alpha\,{\cal O}\bigl((\tfrac{eF}{m^2})^4\bigr)$.
The first (second) line is the result for probe photons polarized in mode $a_1^\mu$ ($a_2^\mu$).
The associated indices of refraction are given by $n_i=1/v_i$. Up to corrections $\sim\alpha\,{\cal O}\bigl((\tfrac{eF}{m^2})^4\bigr)$, their explicit expressions are
\begin{align}
 \left\{\!\!\begin{array}{c}
  n_1 \\ n_2
 \end{array}\!\!\right\}
 =1+\frac{\alpha}{\pi}\frac{1}{90}\!
  \left\{\!\!\begin{array}{c}
  7 \\ 4
 \end{array}\!\!\right\}\!
 \biggl[\Bigl(\frac{eE}{m^2}\Bigr)^2\sin^2\sphericalangle(\vec{k},\vec{E})+\Bigl(\frac{eB}{m^2}\Bigr)^2\sin^2\sphericalangle(\vec{k},\vec{B})-2\frac{eE}{m^2}\frac{eB}{m^2}\,\hat{\vec{k}}\cdot\hat{\vec{s}}\,\biggr]\,.
 \label{eq:nis}
\end{align}

Correspondingly, in generic electromagnetic background fields (for which the expression in the square brackets does not vanish) the two orthogonal probe photon polarization modes $a_1^\mu$ and $a_2^\mu$ propagate with different velocities $v_i\leq1$, or equivalently, experience different indices of refraction, implying that the quantum vacuum subjected to electromagnetic fields acts like a {\it birefringent} medium \cite{Toll:1952,Baier:1967}.

Shining linearly polarized light with overlap to both polarization modes through an electromagnetic field, due to the birefringence effect an ellipticity signal is induced.
The ellipticity signal is typically specified by an angle $\Delta\Phi=2\pi\frac{L}{\lambda}\Delta n$ characterizing the phase shift between the two polarization components, with $\Delta n=n_1-n_2$.
Here $\lambda$ is wavelength of the probe photons and $L$ denotes their optical path length in the electromagnetic field.

As essentially all electromagnetic fields available in the laboratory fulfill $\{\frac{eE}{m^2},\frac{eB}{m^2}\}\ll1$ (cf. above), the vacuum birefringence signal is extremely tiny, making its experimental detection very challenging.
It is so far searched for in experiments using macroscopic magnetic fields of several Tesla and high-finesse cavities to increase the optical path length of the probe photons in the magnetic field \cite{Cantatore:2008zz,Berceau:2011zz,Karbstein:2015qwa}.

An alternative proposal to verify vacuum birefringence with the aid of high-intensity lasers in an all-optical experimental setup has been put forward by \cite{Heinzl:2006xc,Schlenvoigt:2016},
envisioning the combination of an optical high-intensity laser as pump and a linearly polarized x-ray pulse as probe; cf. also \cite{DiPiazza:2006pr,Dinu:2013gaa}.
Resorting to the locally constant field approximation on the level of the effective Lagrangian~\eqref{eq:L_HE}, recently we have reanalyzed this vacuum birefringence scenario \cite{Karbstein:2015xra,Karbstein:2016lby}, rephrasing the phenomenon in terms of a vacuum emission process \cite{Karbstein:2014fva}.

\section{Conclusions}

In these lectures we focused on the quantum vacuum subjected to classical electromagnetic fields.
To this end we explicitly derived the renowned Heisenberg-Euler effective action in constant electromagnetic fields.
As an application, we demonstrated how the experimental signature of vacuum birefringence can be obtained from this effective action.
Let us finally emphasize that the Heisenberg-Euler effective action of course also gives rise to many other signatures of quantum vacuum nonlinearity;
for reviews, see \cite{Dittrich:1985yb,Dittrich:2000zu,Marklund:2008gj,Dunne:2008kc,Heinzl:2008an,DiPiazza:2011tq,Dunne:2012vv,Battesti:2012hf,King:2015tba}.

{\it Acknowledgements}: I would like to thank the organizers of the Helmholtz International Summer School (HISS) - Dubna International Advanced School of Theoretical Physics (DIAS-TH) ``Quantum Field Theory at the Limits: from Strong Fields to Heavy Quarks'' for organizing a very nice and pleasant summer school in Dubna.
Moreover, I am indebted to \!\!
\selectlanguage{russian}
\CYRE\cyrl\cyre\cyrn\cyra\ \CYRM\cyro\cyrs\cyrm\cyra\cyrn
\selectlanguage{english}
for various stimulating discussions \includegraphics[height=2.5mm]{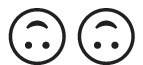}  and to Holger~Gies for many stimulating discussions as well as helpful comments on this manuscript.


\begin{footnotesize}


\end{footnotesize}


\end{document}